# Superconducting FeSe$_{1-x}$Te$_x$ Single Crystals Grown by Optical Zone-Melting Technique

K.W. Yeh[1,]\*, C.T. Ke[1], T.W. Huang[1], T.K. Chen[1], Y.L. Huang[1], P.M. Wu[2], M.K. Wu[1]

[1]Institute of Physics, Academia Sinica, Taipei, Taiwan

[2]Department of Physics, Duke University, Durham, NC, USA

AUTHOR EMAIL ADDRESS: kwyeh@phys.sinica.edu.tw (K.W. Yeh)

\*Corresponding author. Tel: +886-2-27898302; fax: +886-2-27834187, *E-mail address:* kwyeh@phys.sinica.edu.tw (K.W. Yeh)

**ABSTRACT:** A new approach to grow FeSe$_{1-x}$Te$_x$ single crystals with optical zone-melting technique was successfully employed. Crystals with actual composition $0.4 \leq x \leq 1.0$ all show high crystallinity with no phase separation. The ability to visually observe the locally heated melt, and ease of use and control of the image furnace make this method a promising and time-efficient way for obtaining high-quality FeSe$_{1-x}$Te$_x$ crystals. Our results indicate that with adequate heat treatment, the non-uniform distribution of Se and Te atoms in crystal lattice can be effectively eliminated, while the transition width of superconductivity can be reduced to about 2 K, which suggest the crystals are homogeneous in nature.

## 1. Introduction

The family of Fe-based superconductors has grown rapidly, and now comprises of LaOFeAs$_{1-x}$F$_x$ (1111)[1], Ba$_{1-x}$K$_x$Fe$_2$As$_2$ (122)[2], Li$_{1-x}$FeAs (111)[3], and FeSe$_{1-x}$ (11)[4]. FeSe has attracted extensive attention for its simple crystal structure. The FeSe superconducting transition temperature of $T_{c,zero} \sim 8$ K exhibits a compositional dependence, decreasing for both underdoped and overdoped materials[5], as observed in the cuprates. The PbO-type compound FeSe$_{1-x}$Te$_x$ ($x = 0 \sim 1$), where Te substitution has effect on superconductivity, has also been investigated.[6~8] It was found that superconducting transition temperature increases with Te doping, reaching a maximum of $T_{c,onset} \sim 15$ K at about 50 ~ 70 % substitution, and then decreases with more Te doping. For polycrystalline FeSe and FeSe$_{1-x}$Te$_x$ samples, a structural transformation from tetragonal (*P4/nmm*) symmetry into monoclinic lattice (*P112/n*) (or orthorhombic with the defined a-b plane rotated about 45° with respect to the original lattice) with a lower symmetry occurs at around 100 K. The structural distortion, which changes the lattice parameters without breaking magnetic symmetry, was reported and believed to have strong correlation with the occurrence of superconductivity.[4,6] This suggests that a detailed investigation of the magnetic and electronic behavior of these materials and their interplay with structural changes may contribute to a more fundamental understanding of the superconductivity.

In polycrystalline samples, the existence of second phases in grain boundaries projects a considerable degree of uncertainty in the process of analyzing the correlation between the structure and the electronic and superconducting properties. This problem may be overcome by the use of high quality single crystals whose chemical composition and crystal structure can be properly determined. Recently, FeSe$_{1-x}$Te$_x$ crystals with Te doping ($x \geq 0.5$) were grown from the melt using Bridgman method.[9,10] Although resistivity measurements showed superconductivity with onset transition temperatures around 14 K for $x = 0.5 \sim 0.7$, the resistance did not zero down in crystals of $x$ equivalent to 0.9, 0.75, 0.67.[10] Note that in polycrystalline samples with $x = 0.9$, there is also a non-zero residual resistance.[6,7] The fact that the resistance does not completely reach zero indicates a non-uniform distribution of Se and Te, and consequently a low concentration of superconducting component in as-grown crystals. That is why it is important to explore other methods for crystal growth in hopes that more uniform samples can be achieved.

The basic concept of image furnaces is to use ellipsoidal mirrors to focus the light from halogen lamps onto a vertically held rod shaped sample, producing a molten zone, which is then moved along the sample in order to grow a single crystal. The use of light heating makes this technique suitable for both conducting and insulating materials that absorbs infrared easily, which includes the FeSeTe compound under study. In this report, large layered crystals of $FeSe_{1-x}Te_x$ were grown by an optical zone-melting growth method and subsequently homogenized by annealing. Below we present anisotropic magnetic susceptibility and transport measurement results. Importantly, we find that post-annealing temperatures are critical for obtaining uniform and homogeneous single crystal.

## 2. Experimental Methods

Powder materials of Fe (3N purity), Se (3N purity) and Te (5N purity) with desired stoichiometry ($FeSe_{1-x}Te_x$ of $x = 0 \sim 1.0$) were mixed in a ball mill for at least 4 hours. The well-mixed powders were cold-pressed into discs under 400 kg/cm$^2$ uniaxial pressure, and then sealed in an evacuated quartz tube with a pressure less than $10^{-4}$ torr and heat treated at 600 °C for 20 hours. The reacted bulk sample weighing about 5 ~ 8 grams was reground into fine powders and loaded into double quartz ampoule. After being charged, evacuated and sealed, the ampoule was loaded into an optical floating-zone furnace (model 15HD, NEC Nichiden Machinery) with 2 × 1500 W halogen lamps installed as infrared radiation sources. The ampoule was rotated at a rate of 20 rpm and moved at a translation rate of 1 ~ 2 mm/h. To anneal the product, as-grown crystals were heated at a rate of 100 °C/h up to 700 ~800 °C, held at this temperature for 48 hours, cooled at the same rate down to 420 °C and held for another 30 hours, then finally furnace-cooled to room temperature. The grown crystals were stored in a desiccator to avoid decomposition.

Powder XRD measurements were carried out with X-ray diffractometer (Phillips PW3040/60) using Cu $K_\alpha$ radiation, a scanning rate of 0.01° per second, and θ – 2θ scans from 10° to 80°. The thin plate-like crystals were attached onto cupper rings with AB glue and thinned using dual-beam ion milling until electron-transparency. A JEOL JEM-2100F field-emission gun transmission electron microscope was used at 200 keV for imaging, electron diffraction, and energy-dispersive X-ray analysis (EDX) at the thin area of the specimen. Cell parameters, *a*, *c* and gamma angle of a single crystal were collected from the experiment in synchrotron source (BL12b2 at SPring 8) with incident beam of wavelength 0.995 Å. The chemical composition of Fe, Te and Se was determined across the cleaved section parallel to the *ab* plane by energy dispersive X-ray (EDX) equipped in SEM and TEM. DC magnetic susceptibility measurements were performed in a Quantum Design superconducting quantum interference device vibrating sample magnetometer (SQUID VSM). The resistance measurements were carried out in a Quantum Design PPMS system using the standard 4-probe method.

## 3. Results and Discussion

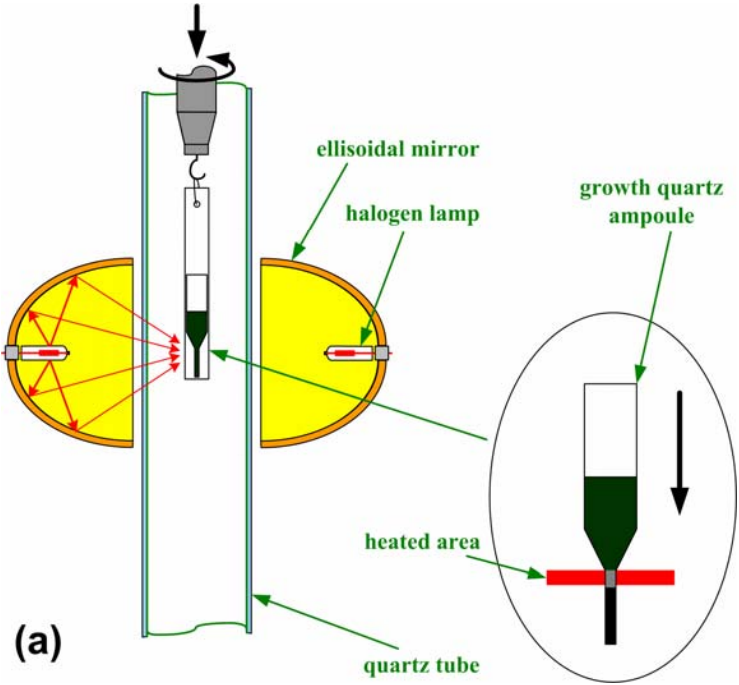

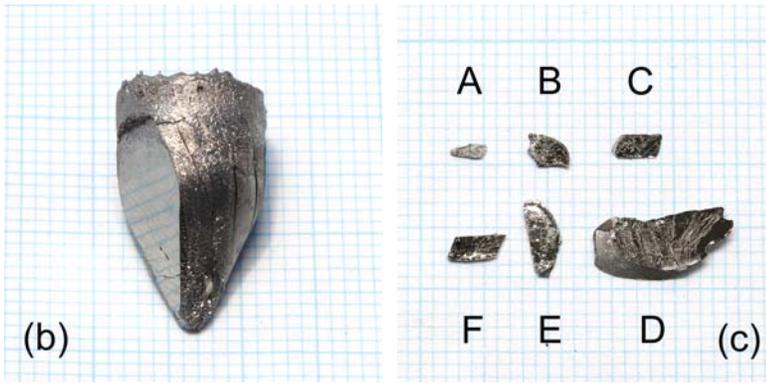

**Figure 1.**

The schematic diagram for optical zone-melting growth is illustrated in Fig. 1(a). Crystallization started from the tip of the quartz ampoule, which moved downward through a 4 mm-width heated area. Subsequently crystals grew along the vertical direction. Figure 1(b) presents a large crystal ingot of FeSe$_{0.3}$Te$_{0.7}$ obtained at a growth rate of 1.5 mm/h. As shown in Fig. 1(c), all of the crystals from $x$ = 0.3 to 1.0 are prone to exfoliation and readily cleaved from an ingot with a razor blade because of the layered property of these tetragonal crystals. It is noteworthy that plate-like crystals of $x$ = 0.3 with dimensions up to 1 × 3 mm$^2$ were successfully obtained by the use of our technique.

Unfortunately, considerable difficulties appeared when growing crystals of low Te concentrations ($x$ < 0.3). Nucleation and grain growth were not successful due to the high viscosity of the melt. Almost polycrystalline chunk was obtained in the growth of FeSe$_{0.8}$Te$_{0.1}$. Evaporations of constituent elements with high vapor pressure, including Se and Te, seem not to be an issue in the case of the local heating of the ampoule. On the other hand, crystallization was successful in the melt of higher Te concentration ($x$ > 0.3) as a result of its lower viscosity.

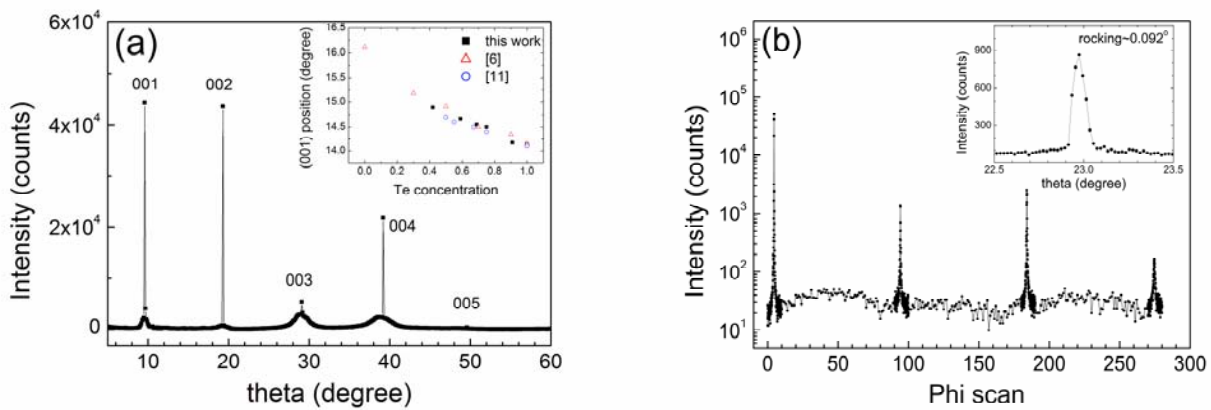

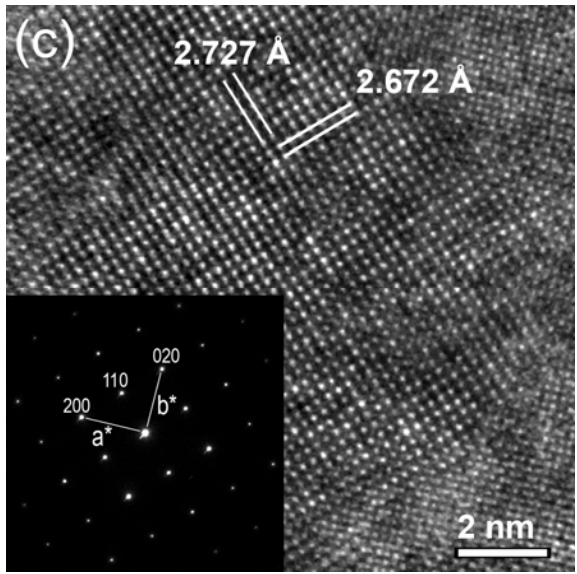

**Figure 2.**

Figure 2(a) shows the typical X-ray diffraction (XRD) pattern on crystal platelet of FeSe$_{0.3}$Te$_{0.7}$. Only (001) peaks were found suggesting that the crystallographic $c$ axis is perpendicular to the plane of the plate-like crystal. The phase purity was examined on crushed crystals by means of powder XRD measurements. All diffraction peaks of crystals with $0.3 < x < 1.0$ belong to the tetragonal phase and no second phase was observed. The phase separation found in polycrystalline samples was not observed within crystals. Typical XRD data give the lattice parameters listed in Table 1. The inset in Fig. 2(a) presents the compositional dependence (Te concentration, $x$) of the (001) peak positions. The peak moves to lower angle as $x$ increases. Figure 2(b) demonstrates four-fold symmetry of the (221) phi scan on FeSe$_{0.3}$Te$_{0.7}$ crystal at room temperature. Only sharp diffraction lines were found, indicating high crystalline quality of the in-plane and out-of-plane structure of the sample. As shown in the inset of Fig. 2(b), the full-width half-maximum (FWHM) is 0.092° in the X-ray rocking curve of (101) Bragg-reflection. EDX analysis of FeSe$_{1-x}$Te$_x$ crystals shown in Table 1 indicates that Te concentrations are close to but always slightly higher than the nominal composition. TEM high resolution image shown in Fig. 2(c) further confirms $z$ axis (001) diffraction pattern of FeSe$_{0.3}$Te$_{0.7}$ crystal and microanalysis in TEM gives the composition FeSe$_{0.25}$Te$_{0.73}$.

Single crystal XRD experiment with synchrotron beam of wavelength 0.995 Å can collect the diffraction of high-index planes revealing detailed structural information in high q range. The result indicates that for FeSe$_{0.3}$Te$_{0.7}$ crystal at room temperature, the crystal lattice already distorts from tetragonal toward a low symmetry structure with space group $P112/n$ ($a$ = 3.8158 Å, $c$ = 6.2300 Å and $\gamma$ = 90.24° at 300 K). This is closely correlated with what was observed for polycrystalline samples.[6] This result indicates that partially substituted Te-atom induces a local strain in the lattice at room temperature. Further measurements on structural distortion at low temperature are needed in order to better understand the origin of superconductivity in this class of materials.

Figure 2(c) shows the high-resolution TEM image and the corresponded electron diffraction of FeSe$_{0.3}$Te$_{0.7}$ crystal. The 001 zone-axis pattern was obtained with only slight tilt from the neutral position, revealing that the plane-normal of the plate surface is the c-axis of the tetragonal lattice. The lattice constants obtained from the electron diffraction pattern were $a$ = 3.819(8) Å and $\gamma$ = 91.21(15)° (lattice constant $c$ is unable to be calculated using this pattern only). The parameter $a$ is in good agreement with the value obtained from synchrotron source, yet the $\gamma$ angle is somewhat larger and inaccurate (with a large standard deviation), which is mainly due to the difficulty in precise determination of the included angle in the diffraction pattern. Nevertheless, the elongated $d_{-110}$ and shrunk $d_{110}$ due to expanded $\gamma$ angle are clearly demonstrated by the atomic-resolution image shown in Fig. 2(c). In addition, the reflections at the (hk0), h+k=2n, h=k=odd positions are strong, while they are expected to be very weak in FeSe and FeTe compounds. This must arise from the deviation of atomic positions caused by inter-substitution of Se and Te.

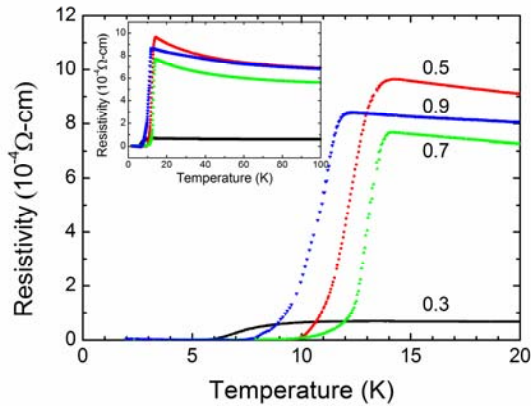

**Figure 3.**

We have measured the in-plane resistivity $\rho_{ab}$ as a function of temperature to characterize transport dynamics. As shown in Fig. 3, FeSe$_{1-x}$Te$_x$ crystals with $0.3 < x < 0.9$ all show a sharp drop in resistivity which indicates the onset of superconductivity. The crystal of $x = 0.3$ has an onset superconducting temperature ($T_{c,onset}$) near 8.9 K. $T_{c,onset}$ of crystals with $0.5 < x < 0.7$ are all near 13 ~ 14 K and crystal with $x = 0.9$ has a $T_{c,onset}$ of 11.5 K. These values agree well with the superconducting "dome" of $T_c(x)$ demonstrated for polycrystalline samples.[6,8] Crystal with $x = 0.9$ where $T_{c,onset}$ is near 11.5 K has a clear $T_{c,zero}$ of 9.2 K. This result is in contrast to the fact that the resistivity does not reach zero for polycrystalline samples of $x = 0.9$ [6,7] and for crystals with $x = 0.67, 0.75, 0.9$ grown by Bridgman method.[10] Therefore, wide transition widths and nonzero resistivity is likely not an intrinsic property, rather a result of the non-uniformity of nonequilibrium sample preparations. It is noteworthy that the superconducting transitions of crystals with all compositions are as narrow as 2 ~ 3 K (as shown in Table 1). Interestingly, the small drop in resistivity near 35 K was not obvious in our crystal with the composition of $x = 0.9$ as reported by Sales et al.[10] The first-order resistive transition occurring near 65 K in FeTe crystal is similar to the reports done by others.

We have found that the use of reacted or unreacted powders as starting materials does not influence the crystallinity and uniformity of crystals; instead, the post-annealing temperature is the crucial variable. The uniformity of FeSe$_{0.5}$Te$_{0.5}$ and FeSe$_{0.3}$Te$_{0.7}$ did not improve much with 700 ~ 780 ºC annealing, but a sharp transition of less than 2.2 K was achieved after 800 ºC treatment. On the other hand, the crystallinity of FeSe$_{0.1}$Te$_{0.9}$ degraded and Te concentration dropped to around 0.7 after 780 ~ 800 ºC annealing. This result suggests the temperature of post annealing should not be higher than 750 ºC for $x = 0.9$. We believe that crystals with higher Te concentrations require lower-temperature annealing probably due to the fact that the melting point decreases with Te concentration. As shown in Table 1, the superconducting transition widths of crystals after annealing were narrowed down to a range of 2 to 2.5 K. The FeSe$_{0.3}$Te$_{0.7}$ sample seems to be the best controlled growth, with a relatively high $T_c$, sharp superconducting transition, good crystallinity, and high phase stability. Our results clearly indicate that the distribution of Se and Te in crystals homogenizes after proper heat treatment.

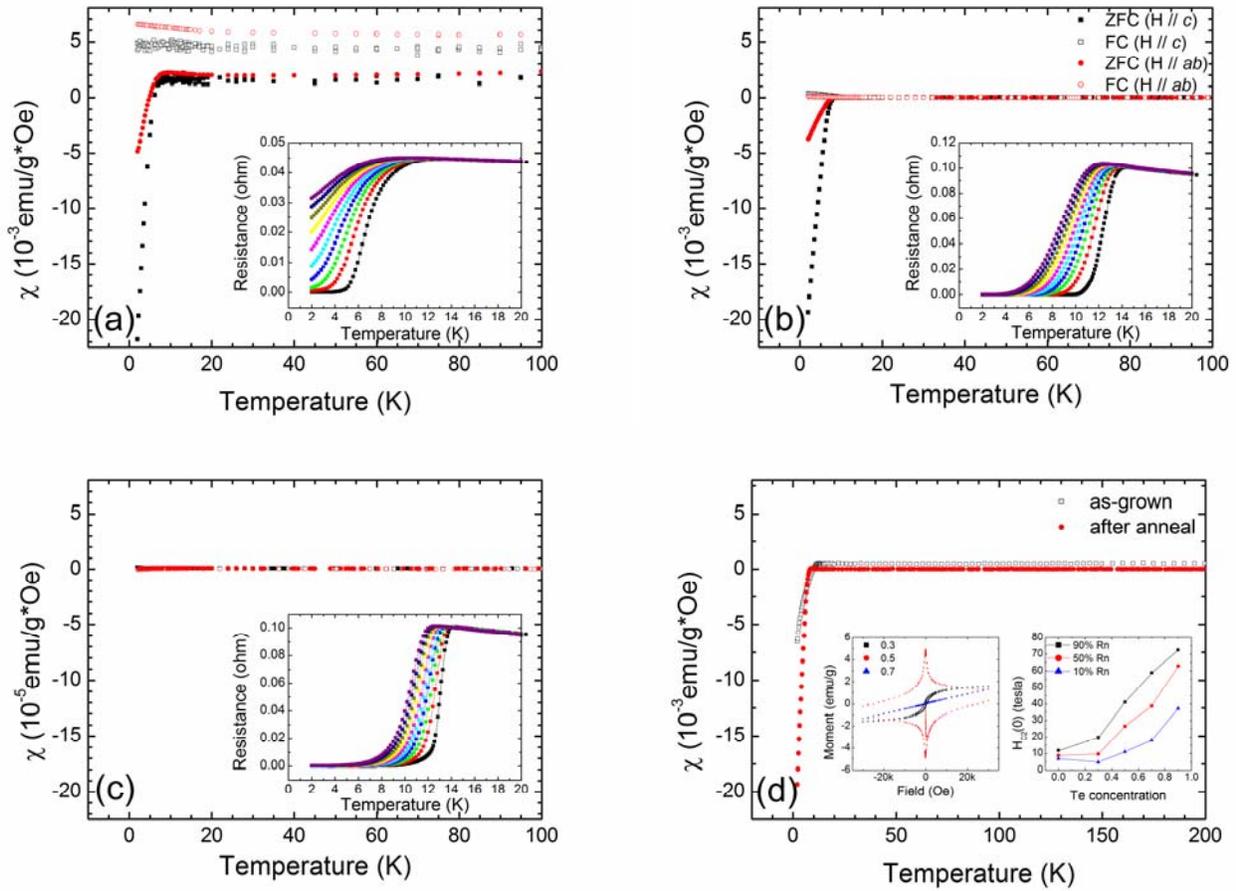

**Figure 4.**

Susceptibility of FeSe$_{1-x}$Te$_x$ single crystals was characterized by a superconducting quantum interference device magnetometer. Field cool (FC) and zero field cool (ZFC) measurements with 30 Oe applied field are presented in Fig. 4. The crystals were oriented with the (001) plane parallel or perpendicular to the magnetic field. The superconducting response in the zero-field cooled mode is more pronounced for the case where the applied field is parallel to crystal $c$ axis, suggesting an anisotropic magnetism and strong 2D characteristics. The diamagnetic response for samples with $x > 0.6$ is small and behaves very much like those of bulk samples[7] indicating low concentration of superconducting component. As seen in Fig. 4(d), lessening the magnitude of Se and Te non-uniform distribution in as-grown crystals by heat treatment can effectively increase the weighting of superconducting component, thereby enlarging the screening response. Compared to bulk samples, smaller Pauli susceptibility in both the ZFC and FC curves in the normal state for high Te-doped samples ($x > 0.5$) indicates a low background of magnetic impurity content (Fe and others). It is seen that the Pauli susceptibility is somewhat larger for $x = 0.3$, which is probably due to lower crystallinity from impurities trapped within the crystal. The extremely small magnetic moment in the field-cooled measurement is possibly a result of the local lattice distortions caused by the substitution with relatively big Te ions, which introduces effective pinning centers.

We plot the field dependence of the resistive transition in the insets of Fig. 4(a) for FeSe$_{0.7}$Te$_{0.3}$ crystal, 4(b) for FeSe$_{0.5}$Te$_{0.5}$ crystal, and 4(c) for FeSe$_{0.3}$Te$_{0.7}$ crystal. The superconducting transition is broadened distinctly in magnetic fields up to 9 T for $x = 0.3$ and is decreasingly broadened as $x$ increases. The upper critical fields, estimated using the WHH formula $H_{c2}(0)=0.69(dH_{c2}/dt)T_c$,[11] was enhanced after annealing. Therefore, the strong impurity scattering effect from the randomly distributed excess Fe$^+$ cannot be ruled out for these high $H_{c2}(0)$ values.[7,12] The right inset of Fig. 4(d) shows that $H_{c2}(0)$ values monotonically increase with increasing Te concentrations after a threshold $x$ of 0.3. It is noteworthy that these values are exceptionally large, comparable to the values found in polycrystalline samples.[8] It further supports the fact that weaker coupling between each plane results in more layer-like transport characteristics as the spacing between neighboring Fe-occupied planes increases with increasing Te concentration up to a threshold concentration, $x = 0.9$. The critical current density at 2 K, estimated from the field dependence of the magnetic moment using Bean's model, shown in the left inset of Fig. 4(d), reaches values on the order of $10^3$ A/cm$^2$ for $x = 0.5$ crystal.

**4. Conclusion**

We proposed a new method for crystal growth such that more uniform samples were achieved. High quality single crystals of FeSe$_{1-x}$Te$_x$ ($0.3 < x < 1.0$) were obtained by optical zone-melting technique. Adequate heat treatment further enhances the uniformity of as-grown crystals. Because of the easy cleavage and high crystallinity of layered FeSe$_{1-x}$Te$_x$ crystals, information of band structures or anisotropic electronic and magnetic properties can be more clearly determined than in lower quality FeSe crystals. The surface-sensitive spectroscopic measurements and X-ray diffraction experiments at low temperature are under way.

**Acknowledgment.** We thank the National Science Council of Taiwan and the US AFOSR/AOARD for their generous financial support and the National Synchrotron Search Centre for experimental support in this work.

**FIGURE CAPTIONS**

**Figure 1.** (a) Schematic diagram of apparatus setup of optical zone-melting method. (b) Single crystal boule of as-grown $FeSe_{0.3}Te_{0.7}$ on a 1 mm grid. The lustrous surface is perpendicular to crystal $c$ axis. (c) The crystal platelets with the (001) face of several mm$^2$ areas. Crystals from A to F represent $FeSe_xTe_{1-x}$ crystals of $x$ = 0.3, 0.5, 0.6, 0.7, 0.9 and 1.0, respectively.

**Figure 2.** (a) X-ray diffraction pattern for $FeSe_{0.3}Te_{0.7}$ crystal. Miller indices for the tetragonal-PbO structure are shown. The (001) peak positions of a series of Te concentrations are plotted in the inset. (b) (221) phi scan on $FeSe_{0.3}Te_{0.7}$ crystal. The inset shows X-ray rocking curve of (101) crystallographic direction for the same crystal. (c) High-resolution TEM image of $FeSe_{0.3}Te_{0.7}$ crystal and the electron diffraction indexed with a tetragonal lattice.

**Figure 3.** Temperature dependence of in-plane resistivity for $FeSe_{1-x}Te_x$ crystals (0.3 < $x$ < 0.9) at temperature range of 0 to 20 K. The inset presents the in-plane resistivity at temperature range of 0 to 100 K.

**Figure 4.** Magnetic susceptibility for (a) $FeSe_{0.7}Te_{0.3}$ crystal, (b) $FeSe_{0.5}Te_{0.5}$ crystal and (c) $FeSe_{0.3}Te_{0.7}$ crystal as a function of temperature with H // $ab$ plane (red circles) and H // $c$ (black squares) in a field of 30 Oe. Solid and hollow symbols represent ZFC and FC signal, respectively. Temperature dependence of resistance under different magnetic fields is plotted in the inset of each plot. (d) Comparison of zero field-cooling signals before and after annealing for $FeSe_{0.5}Te_{0.5}$ crystal. The left inset presents the field dependence of the magnetic moment for $FeSe_{1-x}Te_x$ crystals ($x$ = 0.3, 0.5, 0.7). The right inset demonstrates $H_{c2}(0)$ as function of composition for $FeSe_{1-x}Te_x$ crystals ($x$ = 0, 0.3, 0.5,

0.7, 0.9). $T_c$ is defined by a criterion of 90 % (squares), 50 % (circles) and 10 % (upper triangles) of normal-state resistivity.

**Table 1.** Chemical compositions, superconducting properties and structural characteristics for FeSe$_{1-x}$Te$_x$ single crystals.

| Starting composition | Crystal composition | $T_{c,\text{onset}}$ (K) | $\Delta T_c$ (K) | Out-plane mosaic (º) | In-plane mosaic (º) | Lattice $a$ constant (Å) | Lattice $c$ constant (Å) |
|---|---|---|---|---|---|---|---|
| FeSe$_{0.7}$Te$_{0.3}$ | FeSe$_{0.56}$Te$_{0.41}$ | 8.9 | 3.8 | | | 3.799(1) | 5.942(5) [b] |
| FeSe$_{0.5}$Te$_{0.5}$ | FeSe$_{0.39}$Te$_{0.57}$ | 13.1 | 2.2 | 0.5 | 0.02 | 3.815(4) | 6.031(9) [b] |
| FeSe$_{0.4}$Te$_{0.6}$ | FeSe$_{0.3}$Te$_{0.66}$ | 13.1 | 2.8 | | | 3.827(1) | 6.083(1) [b] |
| FeSe$_{0.3}$Te$_{0.7}$ | FeSe$_{0.25}$Te$_{0.72}$ | 13.6 | 1.5 | 0.2 | 0.01 | 3.8158 | 6.2300 [a] |
| FeSe$_{0.1}$Te$_{0.9}$ | FeSe$_{0.09}$Te$_{0.86}$ | 11.5 | 2.3 | 0.2 | 0.01 | 3.832(6) | 6.255(4) [b] |
| FeTe | FeTe$_{0.91}$ | -- | -- | 0.2 | 0.02 | 3.828(9) | 6.255(4) [b] |

(a) The values were calculated from the experiment in synchrotron source (BL12b2 at SPring 8) with incident beam of wavelength 0.995 Å.

(b) The lattice parameters were estimated by solving the matrix of simultaneous equations (for a tetragonal cell) from low-angle peaks obtained by in-house XRD.

**References**

[1]  Y. Kamihara, T. Watanabe, M. Hirano and H. Hosono, *J. Am. Chem. Soc.* **130**, 3296 (2008).

[2]  M. Rotter, M. Tefel, D. Johrendt, *Phys. Rev. Lett.* **101**, 107006 (2008).

[3]  X.C. Wang, Q.Q. Kiu, Y.X. Lv, W.B. Gao, L.X. Yang, R.C. Yu, F.Y. Li, C.Q. Jin, *Solid State Comm.* **148**, 538 (2008).

[4]  F.C. Hsu, J.Y. Luo, K.W. Yeh, T.K. Chen, T.W. Huang, P.M. Wu, Y.C. Lee, Y.L. Huang, Y.Y. Chu, D.C. Yan, M.K. Wu, *PNAS* **105**, 14262 (2008).

[5]  T.M. McQueen, Q. Huang, V. Ksenofontov, C. Felser, Q. Xu, H. Zandbergen, Y.S. Hor, J. Allred, A.J. Williams, D. Qu, J. Checkelsky, N.P. Ong, R.J. Cava, *Phys. Rev. B* **79**, 014522 (2009).

[6]  K.W. Yeh, T.W. Huang, Y.L. Huang, T.K. Chen, F.C. Hsu, Wu, Phillip M., Y.C. Lee, Y.Y. Chu, C.L. Chen, J.Y. Luo, D.C. Yan, M.K. Wu, *Euro Physic Letter*, **84**, 37002 (2008).

[7]  M.H. Fang, H.M. Pham, B. Qian, T.J. Liu, E.K. Vehstedt, Y. Liu, L. Spinu, Z.Q. Mao, *Phys. Rev. B* **78**, 224503 (2008).

[8]  K.W. Yeh, H.C. Hsu, T.W. Huang, P.M. Wu, Y.L. Huang, T.K. Chen, J.Y. Luo, M.K. Wu, *J. Phys. Soc. Jpn*. **77**, 19 (2008).

[9]  G.F. Chen, Z.G. Chen, J. Dong, W.Z. Hu, G. Li, X.D. Zhang, P. Zheng, J.L. Luo, N. L. Wang, *Phys. Rev. B* **79**, 140509 (2009).

[10]  B.C. Sales, A.S. Sefat, M.A. McGuire, R.Y. Jin, Y.D. Mandrus, *Phys. Rev. B* **79**, 094521 (2009).

[11]  N.R. Werthamer, E. Helfand, P.C. Hohenberg, *Phys. Rev.* **147**, 295 (1966).

[12]  L.J Zhang, D.J. Singh, M.H. Du, *Phys. Rev. B* **79**, 012506 (2009).